\documentclass{moriond}

\usepackage[symbol]{footmisc}
\usepackage[english]{babel}
\usepackage{graphicx}
\usepackage{graphics}
\usepackage{braket}
\usepackage{bbold}
\usepackage{amssymb}
\usepackage{amsmath}
\usepackage{nicefrac}
\usepackage{dcolumn}
\usepackage{bm}
\usepackage{slashed}
\usepackage{datetime}
\usepackage{mciteplus}
\usepackage{multirow}
\usepackage{bbold}
\usepackage{siunitx}
\usepackage{booktabs}
\usepackage{color, soul}
\usepackage[usenames,dvipsnames]{xcolor}
\usepackage{float}
\usepackage[utf8]{inputenc}
\usepackage[normalem]{ulem}
\usepackage{mathtools}

\newcommand{\LQCD}{\Lambda_{\rm QCD}}
\newcommand{\MSb}{\overline{\rm MS}}

\newcommand{\DY}{\Delta Y}

\def\be{\begin{equation}}
\def\ee{\end{equation}}
\def\bea{\begin{eqnarray}}
\def\eea{\end{eqnarray}}

\newcommand{\Photo}{\includegraphics[height=35mm]{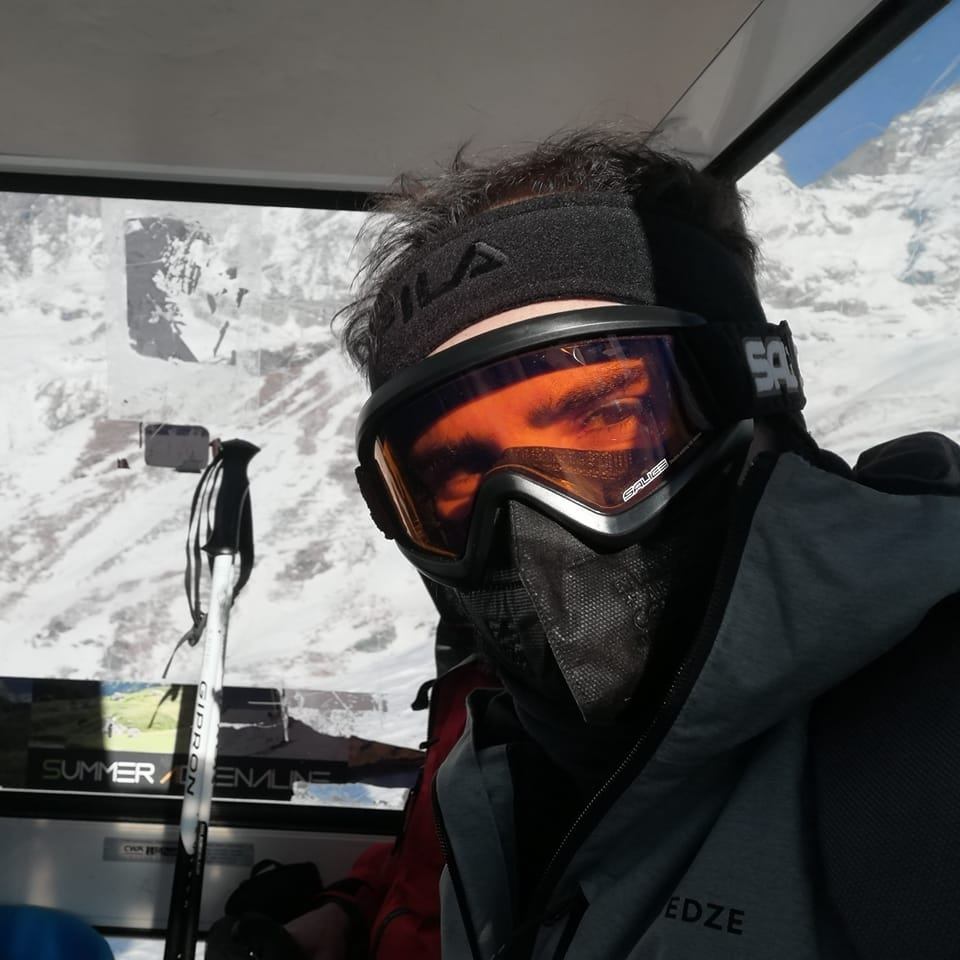}}

\begin{document}
\vspace*{4cm}
\title{High-energy resummed Higgs-plus-jet distributions \\ at NLL/NLO* with {\tt POWHEG+JETHAD}}

\author{F.G. Celiberto,$^{\;1,2\;}$\footnote[2]{Corresponding Author. Electronic address: \href{mailto:francesco.celiberto@uah.es}{francesco.celiberto@uah.es}} L. Delle Rose,$^{\;3,4}$ G. Gatto,$^{\;3,4}$ M. Fucilla$^{\;3,4,5}$ and A. Papa$^{\;3,4}$}

\address{
${}^1$INFN-TIFPA Trento Institute of Fundamental Physics and Applications, I-38123 Povo, Trento, Italy
\\
${}^2$Universidad de Alcal\'a (UAH), E-28805 Alcal\'a de Henares, Madrid, Spain
\\
${}^3$Dipartimento di Fisica, Universit\`a della Calabria, I-87036 Arcavacata di Rende, Cosenza, Italy
\\
${}^4$INFN-Cosenza, I-87036 Arcavacata di Rende, Cosenza, Italy
\\
${}^5$Universit\'e Paris-Saclay, CNRS, IJCLab, 91405 Orsay, France
}

\maketitle\abstracts{
We study the inclusive production, at hadron colliders, of a Higgs boson and a jet widely separated in rapidity.
Kinematic sectors explored fall into the so-called semi-hard regime, where both fixed-order and high-energy dynamics come into play.
Therefore, we propose a first version of a matching procedure aimed at combining NLO fixed-order computations, as obtained from {\tt POWHEG}, with the NLL resummation of energy logarithms via {\tt JETHAD}.
According to our knowledge, this represents a novel implementation of a matching in the context of the high-energy resummation at NLL and for rapidity-separated two-particle final states.}

\section{Hors d'{\oe}uvre}
\label{sec:introduction}


Since the discovery of the Higgs boson at the LHC, a great effort has been made both from a theory and an experimental viewpoint to perform stringent tests of the Standard Model, as well as to hunt for New Physics.
From a perturbative Quantum Chromodynamics (QCD) perspective, accurate benchmarks of the gluon-fusion channel are needed.~\cite{Dawson:1990zj,Djouadi:1991tka}
On one hand, higher-order calculations are core elements of the well established \emph{collinear factorization}, where physical cross sections take the form of a one-dimensional convolution between on-shell coefficient functions and collinear parton distribution functions (PDFs).
On the other hand, modern colliders allows us to enter kinematic sectors where one or more classes of logarithms, naively neglected by collinear factorization, emerge. Those logarithms can be large enough to hamper the convergence of the perturbative series,
thus calling for the use of adequate all-order techniques, known as \emph{resummations}. 
We turn our attention to the \emph{semi-hard} domain.~\cite{Gribov:1983ivg,Celiberto:2017ius}
Here, a rigorous scale hierarchy, $\LQCD^2 \ll \mu_i^2 \ll s$ ($\LQCD$ is the QCD scale, $\mu_i$ are process-typical hard scales, and $s$ is the squared center-of-mass energy), leads to the rise of large energy logarithms, $\ln(s/\mu_i)$.
The most powerful mechanism to account for these logarithms is the Balitsky--Fadin--Kuraev--Lipatov (BFKL) resummation,~\cite{Fadin:1975cb,Balitsky:1978ic} available both in the leading-logarithmic (LL) and in the next-to-leading logarithmic (NLL) approximation (see~\cite{Caola:2021izf}$^-$\cite{Fadin:2023roz} for recent next-to-NLL advancements).
Notably, the BFKL approach can improve our way to unveil the proton content at low-$x$.~\cite{Ball:2017otu}$^-$\cite{Bolognino:2021niq}
Golden channels to access the semi-hard regime are hadroproductions of two particles featuring high transverse masses (acting as $\mu_i$ scales) and separated by a large rapidity distance, $\DY$.
The last condition enhances undetected gluons strongly ordered in rapidity, thus generating large logarithmic corrections to be resummed by BFKL.
To get a reliable description of these reactions at LHC energies, a \emph{hybrid high-energy and collinear factorization} (HyF) was built up.~\cite{Colferai:2010wu}$^-$\cite{Bolognino:2021mrc}
We refer to~\cite{Deak:2009xt}$^-$\cite{Silvetti:2022hyc} for similar studies, close in spirit to ours.
HyF partonic cross sections are elegantly factorized as convolutions between the process-universal BFKL Green's function, known within the NLL, and two singly off-shell coefficient functions. 
Then, they are convoluted with standard collinear PDFs.
The HyF global accuracy can be pushed up to NLL/NLO, provided that coefficient functions are known at NLO. 
Otherwise, one must rely upon a partial next-to-leading accuracy, labeled as NLL/NLO*.
Semi-hard reactions, which have been studied within a full NLL/NLO accuracy, are emissions of two Mueller--Navelet jets~\cite{Ducloue:2013bva}$^-$\cite{Celiberto:2022gji}, light-flavored hadrons~\cite{Celiberto:2017ptm}$^-$\cite{Celiberto:2022kxx}, or heavy-flavored ones.~\cite{Celiberto:2021dzy}$^-$\cite{Celiberto:2022keu}
In this study we focus on the semi-inclusive Higgs-plus-jet process, whose description was afforded at NNLO in perturbative QCD~\cite{Boughezal:2013uia}$^-$\cite{Boughezal:2015dra} and within the next-to-NLL in the \emph{transverse-momentum resummation} formalism.~\cite{Monni:2019yyr}
When, however, the final-state rapidity increases, the weight of energy logarithms grows.
Therefore, the high-energy resummation from our hybrid factorization can serve as a useful tool to improve the description of Higgs-plus-jet differential distributions.~\cite{Celiberto:2020tmb,Celiberto:2023rtu,DelDuca:1993ga}
We present the {\tt POWHEG+JETHAD} method, a first, prototype version of a matching procedure aimed at combining NLO fixed-order computations with the NLL resummation for Higgs-plus-jet rapidity and transverse-momenta spectra.

\section{Higgs-plus-jet distributions
at NLL/NLO*}
\label{sec:matching}

On one side, from recent HyF predictions for the Higgs $p_H$-spectrum in semi-hard Higgs-plus-jet emissions at the LHC, a fair stability under higher-order corrections as well as energy-scale variations came out.
On the other side, very large deviations from the fixed-order background emerged. Their size reaches almost two orders of magnitude when $p_H \gtrsim 120$~GeV.~\cite{Celiberto:2020tmb}
An analogous pattern was found for the $\DY$-spectrum at LHC and FCC energies.~\cite{Celiberto:2023rtu}
Thus, with the aim of catching precision-level corrections from our resummation on top of the standard collinear factorization, we have built a first and novel matching procedure,\footnote{Another scheme was recently proposed to match NLO quarkonium cross sections with the LL resummation.~\cite{Lansberg:2021vie}} based on a numeric implementation of exact removal of the corresponding double counting at the NLL*.
More in particular, since NLO corrections to the Higgs singly off-shell coefficient function were computed only recently~\cite{Celiberto:2022fgx,Hentschinski:2020tbi} and they have not yet been implemented in {\tt JETHAD}, our reference technology,~\cite{Celiberto:2020wpk,Celiberto:2022rfj} we rely on a NLL/NLO* treatment.
Equation~\eqref{eq:matching} schematically depicts our matching

\begin{equation}
\label{eq:matching}
\begin{split}
 \underbrace{{\rm d}\sigma^{\rm NLL/NLO\text{*}}(\Delta Y, \varphi, s)}_{\text{\colorbox{OliveGreen}{\textbf{\textcolor{white}{NLL/NLO*}}} {\tt POWHEG+JETHAD}}} 
 = 
 \underbrace{{\rm d}\sigma^{\rm NLO}(\Delta Y, \varphi, s)}_{\text{\colorbox{gray}{\textcolor{white}{\textbf{NLO}}} {\tt POWHEG} w/o PS}}
 +\; 
 \underbrace{\underbrace{{\rm d}\sigma^{\rm NLL\text{*}}(\Delta Y, \varphi, s)}_{\text{\colorbox{red}{\textbf{\textcolor{white}{NLL* resum}}} (HyF)}}
 \;-\; 
 \underbrace{\Delta{\rm d}\sigma^{\rm NLL/NLO\text{*}}(\Delta Y, \varphi, s)}_{\text{\colorbox{white}{\textbf{NLL* expanded}} \!\!at NLO}}}_{\text{\colorbox{NavyBlue}{\textbf{\textcolor{white}{NLL*}}} {\tt JETHAD} w/o NLO* double counting}}
 \,.
\end{split}
\end{equation}

\begin{figure*}[!t]
\centering

   \includegraphics[scale=0.38,clip]{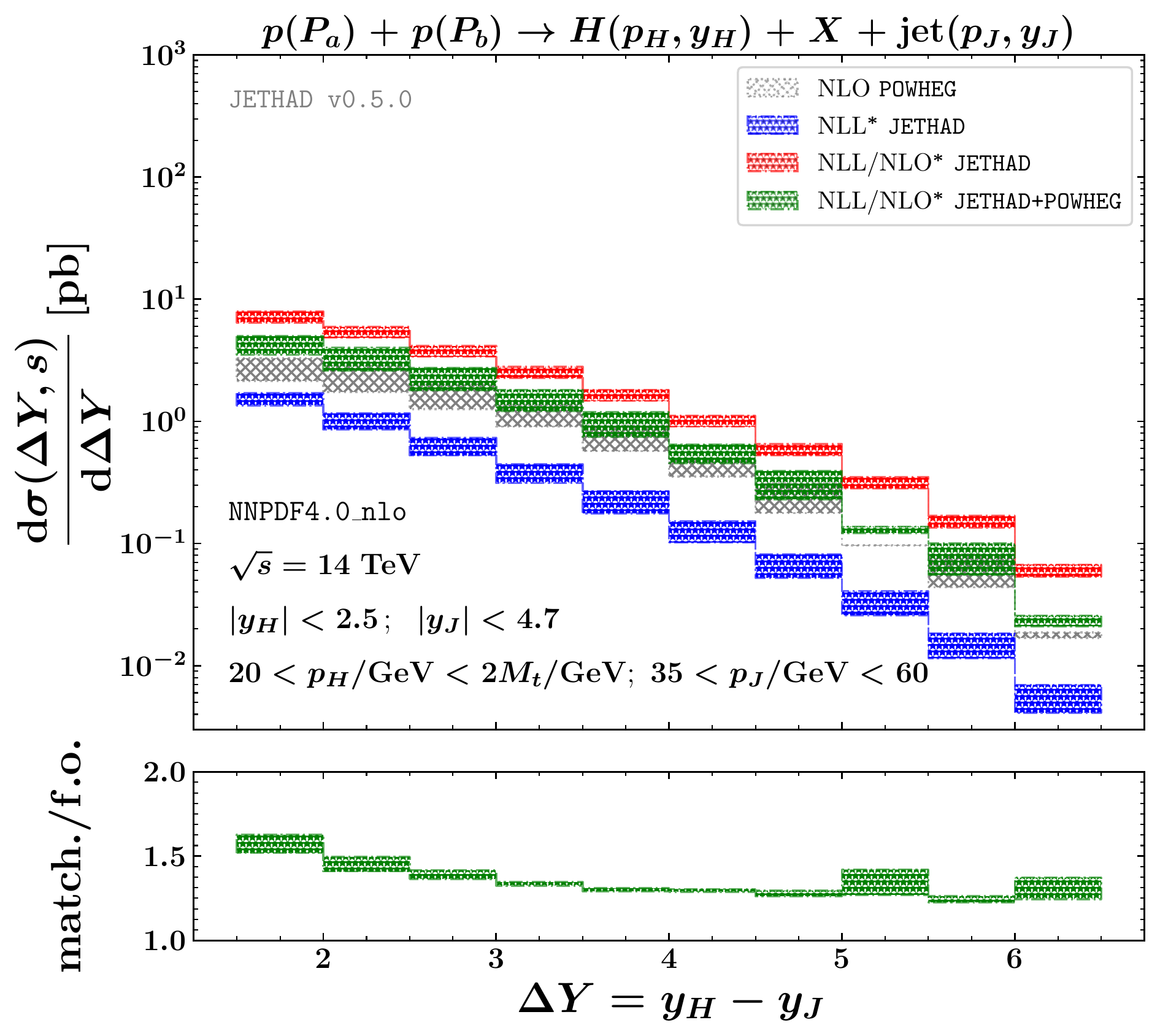}
   \hspace{0.30cm}
   \includegraphics[scale=0.38,clip]{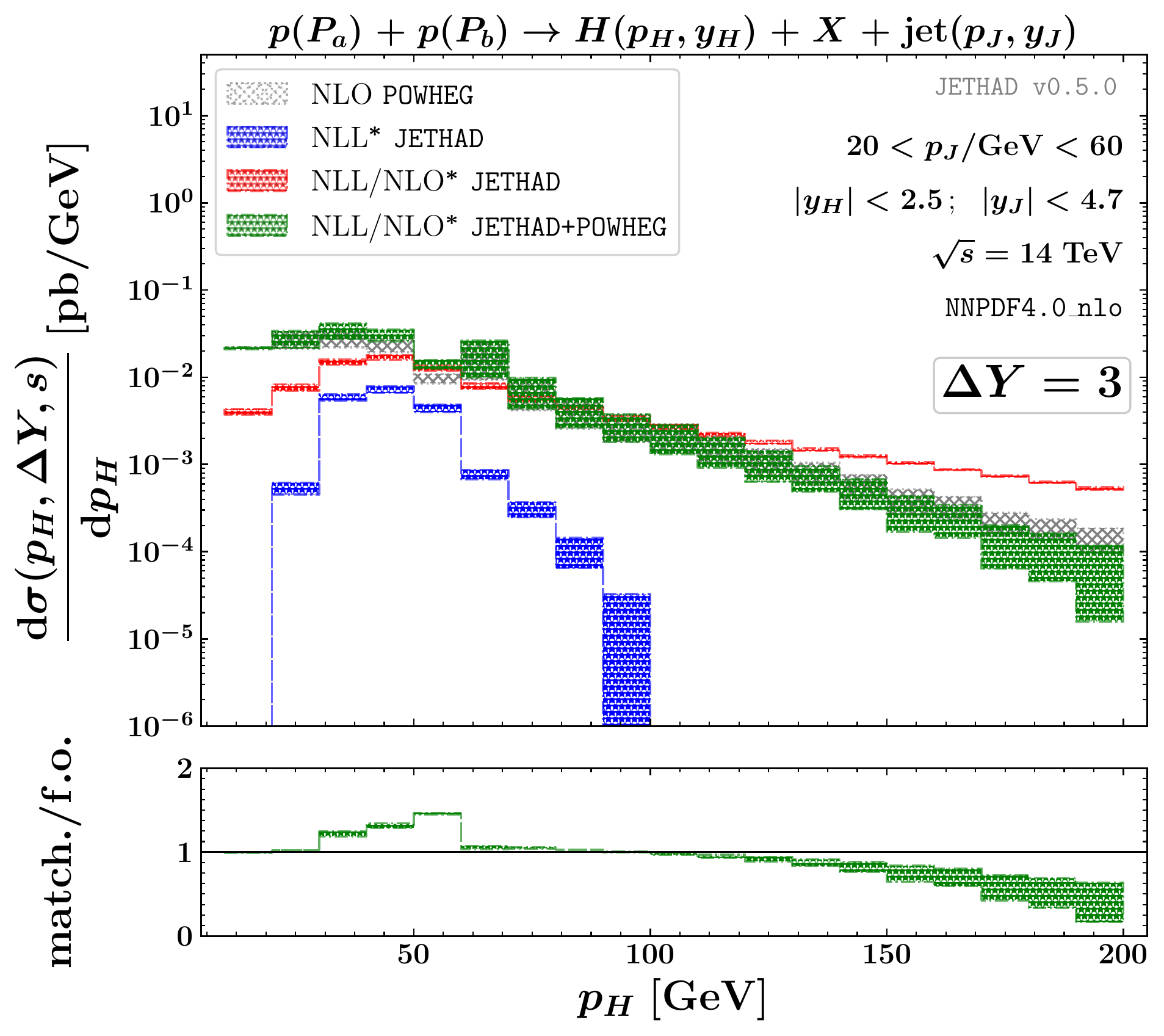}

\caption{Higgs-plus-jet rapidity-interval (left) and transverse-momentum (right) spectrum at 14~TeV LHC.
Shaded bands embody the effect of $\mu_{R,F}$ variations in the $1 < C_{\mu} < 2$ range. Text boxes are for kinematic cuts.
}

\label{fig:spectrum}
\end{figure*}

Any differential cross section, matched at NLL/NLO* (green; for the sake of example, given in Eq.~\eqref{eq:matching} as a function of $\DY$, azimuthal-angle distance, and energy), is written as a sum of the NLO fixed-order term (gray) from {\tt POWHEG}~\cite{Nason:2004rx}$^-$\cite{Hamilton:2012rf} and the NLL* signal (blue) from {\tt JETHAD}. The latter is the HyF-genuine NLL* resummed term (red) minus the NLL* expanded at NLO, namely without the double-counted term.
Removing it inside {\tt JETHAD} instead of {\tt POWHEG} has the double advantage of $(i)$ making our technology dynamically compatible with different matching procedures, and $(ii)$ eliminating spurious power-correction contaminations naturally accounted for by HyF to all orders.
{\tt POWHEG} is used for purely fixed-order predictions, namely without considering \emph{parton-shower} (PS) effects.~\cite{Alioli:2022dkj}$^-$\cite{vanBeekveld:2022ukn}
In Fig.~\ref{fig:spectrum} we present preliminary results for the rapidity-distance (left) and transverse-momentum (right) spectra at 14~TeV.
Calculations were done in the $\MSb$ renormalization scheme. Accordingly, {\tt NNPDF4.0} PDFs at NLO were employed.~\cite{NNPDF:2021njg}
Color code matches the one adopted in Eq.~\eqref{eq:matching}. Ancillary panels below plots show that the high-energy signal is present and its correction to the fixed-order background is up to 50\%.

\section{From HyF to NLL/NLO: A path toward precision}
\label{sec:conclusions}

We have presented a novel and prototype version of a matching procedure, relying on the {\tt POWHEG}~\cite{Nason:2004rx}$^-$\cite{Hamilton:2012rf} and {\tt JETHAD}~\cite{Celiberto:2020wpk,Celiberto:2022rfj} technologies. It is aimed at combining NLO fixed-order computations with the NLL resummation of energy logarithms.
We plan to: $(i)$ assess the weight of full NLO contributions,~\cite{Celiberto:2022fgx,Hentschinski:2020tbi} $(ii)$ gauge the impact of heavy-quark finite-mass corrections,~\cite{Jones:2018hbb,Bonciani:2022jmb} $(iii)$ compare our results with the ones obtained via the PS~\cite{Alioli:2022dkj}$^-$\cite{vanBeekveld:2022ukn} and the {\tt HEJ}~\cite{Andersen:2022zte,Andersen:2023kuj} approaches.

\section*{Acknowledgments}
 
This work is supported by the Atracci\'on de Talento Grant n. 2022-T1/TIC-24176, Madrid, Spain, and by the INFN\-/QFT\-@COL\-LIDERS Project, Italy.


\end{document}